# Aurora Candidates from the Chronicle of *Qīng* Dynasty in Several Degrees of Relevance

Akito D. Kawamura (1), Hisashi Hayakawa (2), Harufumi Tamazawa (1), Hiroko Miyahara (3), Hiroaki Isobe (4, 5)

(1) Kwasan Observatory, Kyoto University, Japan

(2) Graduate School of Letters, Kyoto University, Japan

(3) Musashino Art University, Tokyo, Japan

(4) Graduate School of Advanced Integrated Studies in Human Sustainability, Kyoto University, Japan

(5) Unit of Synergetic Studies for Space, Kyoto University, Japan

## Abstract

We present the result of the survey of sunspots and auroras in *Qīngshǐgǎo* (清史稿), the draft chronicle of *Qīng* dynasty, for the period of 1559-1912 CE, as a sequel of the series of works surveying historical sunspot and aurora records, and providing online data to the scientific community regarding the attained results. In total of this *Qīngshǐgǎo* survey, we found 111 records of night-sky luminous events with the keywords such as vapor (氣, *qì*), cloud (雲, *yún*), and light (光, *guāng*), which may indicate auroras as well as some other phenomena. Similarly keyword survey for sunspots were done, but no sunspot record was found. In comparison with the aurora records in the western world, we found 14 of the 111 records have a corresponding record of simultaneous observation in the western world and hence are very likely to be aurora. In order to investigate the likeliness of the rest of the record being aurora, we calculated the lunar age and the phase of a solar cycle for each record. After these calculations, notable fraction of these records clustered near the full moon were found statistically doubtful in considerations with atmospheric optics, meanwhile a few records near the new moon could be more likely interpreted as auroras including three records during the Maunder minimum.

## 1. Introduction

Solar activity in the past is of great interest from the viewpoint of the long-term variation of solar dynamo as well as the occurrence probability of rare and extreme events. Especially the latter has been attracting the interests recently because of the observational discoveries of "superflares" in other Sun-like (G type) stars (Schaefer et al. 2000; Maehara et al. 2012; Shibayama et al. 2013; Schrijver et al. 2012; Notsu et al. 2015a, 2015b) as well as the footprints of high cosmic-ray radiation events in wood rings that may be related to large solar flares (Miyake et al. 2012, 2013). To understand long-term solar activity, one could trace it in historical documents by the record of auroras in low latitude resulting from severe solar activity, whose importance was also stated in previous studies (i.e. Eddy 1980). Such solar studies would heavily depend on surveys for illuminating phenomena in the night sky from historical documents from China (Schove & Ho (1967); Keimatsu 1970-1976; Yau & Stephenson 1988; Saito & Ozawa 1992; Yau et al. 1995), Japan (Kanda 1933; Matsushita 1956; Nakazawa et al. 2004; Shiokawa et al. 2005), Korea (Lee et al. 2004), the Arab nations (Vaquero & Gallego 2001, Basurah 2006), Russia (Vyssotsky 1949), Europe (Fritz 1873; Link 1962; Dall'Olmo 1979; Stothers 1979; Vaquero & Trigo 2005; Vaquero et al. 2010), North America (Broughton 2002), and the Tropical Atlantic Ocean (Vázquez & Vaquero 2010). We also have published the lists of aurora candidates included in the Chinese chronicles of *Sòng* (宋) dynasty (Hayakawa et al. 2015).

In this paper, we focus on the *Qīng* dynasty (1559 - 1912 CE[1], from the birth of Nurhaci, the founder of Manchurian dynasty preceding, to *Qīng* dynasty to the end of *Qīng* dynasty) by using *Qīngshǐgǎo* (清史稿) as our source document, which is the draft chronicle of *Qīng* dynasty covering from the rise of *Qīng* dynasty in 1616 CE to the foundation of Republic of China in 1912 CE. We provide the table of night-sky luminous events from *Qīngshǐgǎo* with the calculated lunar age, and discuss their statistical properties and possible interpretations for other physical phenomena. No sunspot record was found in *Qīngshǐgǎo* as we explain later.

---

1 CE: "Common Era" is nonreligious alternative of AD.

A notable feature is the chronological position of *Qīng*, which is after the start of the telescopic observation of the sun in the 17th century, and hence the group sunspot number (GSN), a reliable indicator of the yearly-based sunspot activity, is available at Sunspot Index and Long-term Solar Observations (SILSO; website http://www.sidc.be/silso/) in several versions; the original version (Hoyt & Schatten 1998a, 1998b) and alternate series (version 2.0) with "backbone" method (Svalgaard & Schatten 2016), "active-days" method (Usoskin et al 2016), and "error-testing" method (Cliver & Ling 2016). For this paper we adopt the "backbone" series of GSN (for more historical background of this series, Clette et al. 2014 is recommended). However, it should be noted that even without sunspots the eruption of a quiescent prominence could produce large magnetic storm (McAllister et al. 1996). It is interesting if there is any record of low-latitude aurora during the Maunder minimum (around 1645 - 1715 CE) and Dalton minimum (around 1790 - 1830 CE), both periods are before the first observation of flare, as known as Carrington event of Sep. 1, 1859 CE (Carrington 1859) .

# 2. Method

## 2.1 Source documents

To execute mechanical surveys of keywords on a Chinese document, we use Scripta Sinica, (http://hanchi.ihp.cinica.edu.tw) provided by Academia Sinica in Taiwan (http://www.sinica.edu.tw/) as similar to our previous paper (Hayakawa et al. 2015) but with the different source document,

*Qīngshĭgăo* (清史稿, Zhào 1976), the draft chronicle of *Qīng* dynasty. According to the Chinese tradition, an official history (正史) of a dynasty must be compiled and published by the succeeding dynasty as a confirmation of a "legitimate successor." *Qīngshĭgăo* was edited for this purpose, but due to the political situation after the fall of *Qīng,* the role of *Qīngshĭgăo* as the official historical document of China became controversial (Chen 2004). Anyway we focused on *Qīngshĭgăo* because Scripta Sinica adopts it even if it is not a official history.

In order to survey candidates qualifying as aurora observations, we searched the following keywords: vapor (氣, *qì*), cloud (雲, *yún*), and light (光, *guāng*). Once the passages including the keywords were picked up, we removed those explicitly associated with the sun or moon, or stated them as happening during daytime. We also searched black spots (黒子, *hēizi*) and black vapors (黒氣, *hēiqì*) for sunspots, but we did not find any record in *Qīngshĭgăo* . Therefore, we discuss only the records for seeking aurora candidates in the rest of this paper. The result of this keyword survey is presented in Table 1.

There exist possibilities of errors which might have been generated while copying handwritten documents during the compilation of a document. This kind of error is hard to find and to fix without the access to the original text. Only notable case is, with the record written with non-existing date, found error at the date conversion to western calendar. However, the relevance of date and record context with considerations of possible copying errors must be discussed among literature and history specialists. Therefore, we use texts as they are published and treat a record with the non-existing date as an error-dated record.

## 2.2 Observation site

In Chinese dynasties, astronomical observations were mostly performed in the capital city. As Keimatsu (1976) pointed out, after 1368 CE *Míng* dynasty started observations at new observatories in regional cities, and *Qīng* dynasty succeeded not only the central observatory in *Běijīng* (*Qīngshǐgǎo* Astronomy II, p1035) but also these regional observatories. When an observation site is specified in a record, we include it in our aurora candidate list shown in the table 1. When the observation site was not explicitly stated, the observation was likely performed in the capital city, *Běijīng* (北京).

Although we could not find a comprehensive catalogue of observatories in *Qīngshǐgǎo*, we found 53 regional sites associated with records in *Qīngshǐgǎo*, widely spread from 23º to 42º North in latitude and 89º to 122º East in longitude. Many of these places were located in the *Héběi* (河北) and *Shāndōng* (山東) provinces. The cities and corresponding geographical positions are listed in Table 2.

## 2.3 Lunar age

It is unlikely that ancient Chinese observers knew the physical nature of auroras, and hence they used non-scientific terms to express the faint light seen in the sky. Therefore, it is inevitable that other natural phenomena, such as lunar halos, were included in the list after the literal elimination of those clearly written with sun and moon or during the daytime. In the case of atmospheric optical events involving the moon (such as lunar halo) contaminating our survey result, lunar age calculation may reveal the significance of contamination. It is because observations of atmospheric optical phenomena and auroras would depend differently on the presence of the moon; atmospheric optical phenomena with the moon as their light source should appear when the moon is bright, while auroras are more easily seen in the absence of the moon, although strong auroras are still visible in full moon.

In order to calculate the lunar age (moon phase) of each record, we employed “moonphase.pro” and associated algorithms from the IDL library of NASA Goddard Space Flight Center for lunar luminosity calculation, based on Meeus (1998). We also developed an IDL code to find local luminosity minimums around the approached dates, concerning the previous and next new moon occurrences. Our minimum finding algorithm was a combination of Newton’s method and bisection search, guaranteeing a local minimum of luminosity for ±1 second. Finally, we calculated the lunar age of any given date normalized for the period between the new moons. Therefore, the lunar age is considered 0.0 for the new moon and 1.0 for the next new moon, with blank for incomplete date information.

**2.4 Phase of a solar cycle**

Other interesting characteristics of records are the relation with phase of solar cycles because aurora may indicate a solar cycle. Solar cycles are determined based on the "backbone" series of version 2 GSN (Svalgaard & Schatten 2016) and studies on radiocarbon in tree rings for the period of Maunder minimum where the cycles are found to be starting at CE 1633, 1644, 1658, 1672, 1686, and 1700 (Miyahara et al. 2004; Miyahara et al. 2008; Miyahara 2010). We express the phase of solar cycles in the normalized values of between 0.0 and 1.0, start and end of a solar cycle respectively.

# 3. Results and Discussion

## 3.1 Overall result

After the keyword search and the filtration described in the previous section, we retrieved records of aurora candidates from *Qīngshĭgăo*, which are listed in table 1. Quality of records varies in date information; three of them have no or error date information, 40 of them have incomplete date information but at least with year, 68 of them have complete date information.

Color of aurora also varies in records; 77 records are white, 11 records are red, and 23 records are described as other colors or combinations of multiple colors. In this section, we will investigate these records with Western observations and lunar age analysis in order to discuss the relevance of these candidates to aurora. The details of the each record with the original text and the bibliographic information are available at our website (http://www.kwasan.kyoto-u.ac.jp/~palaeo/).

### 3.2 Comet-like records with long durations

Some records in *Qīngshĭgăo* state long duration phenomena, from several days to more than one month, which may be regarded as comets. Due to its nature, a low latitude aurora is unlikely to remain visible for several days or more unless there are strong coronal mass ejections (CMEs) successively emitted to the Earth, injecting enormous amounts of magnetic flux into the Earth's magnetosphere. Although this scenario of successive CMEs cannot be ruled out, appearance of a comet provide an alternative, apparently more reasonable interpretation. Indeed, there are records in the western world of the great comet in 1668 CE and Kirch's comets in late 1680 CE. The records ID27-34 in Table 1, dated from February to March in 1668, are found with simultaneous observation of the comet (March 3 and 9 at Lisbon, Portugal and Bologna, Italy, respectively, according to Lynn 1888). Similarly, the records ID 50-56 coincide with the Western observations of Kirch's comet, such as those described by Yeomans (1991).

In order to select the records that are of likely comets, we took two steps to filter out records. First we picked up those records that describe explicitly the duration of the event longer than 4 days as the comet candidates. Then those records that are dated within the duration of the likely comets are also judged as the comet candidates. As a result, 26 records were classified as comet candidates; which are IDs 13, 14, 15, 27, 28, 29, 30, 31, 32, 33, 34, 49, 50, 51, 52, 53, 54, 55, 56, 58, 59, 60, 62, 94, 95, 98, and 106. Among these records, ID 55 is the only record of blue-white, and the color of all others is white. These 28 comet candidates will be excluded from the following analyses.

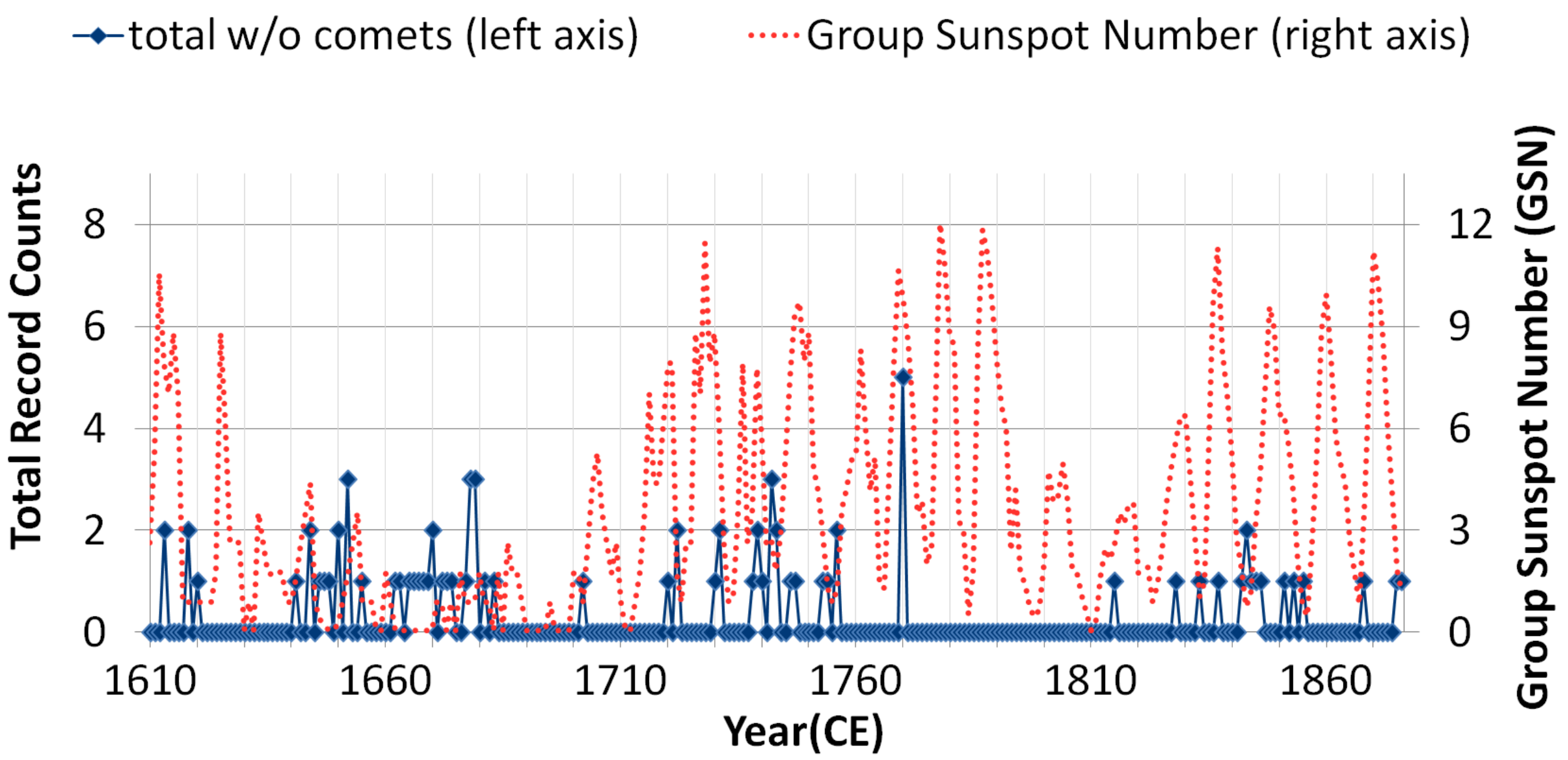

Figure 1. Yearly-based total record counts of night-sky luminous events excluding comet-like records from *Qīngshĭgăo* (blue line and diamond) with group sunspot number (GSN, red dotted line) displayed over the time line. (Color online)

### 3.3 Chronological comparison with the group sunspot number

Since the period of Qing follows the beginning of telescopic observation, we were able to compare our results with the historical records of group sunspot numbers (GSN). Figure 1 shows the yearly-total record counts of night-sky luminous events excluding comet candidates and the "backbone" series of GSN (Svalgaard & Schatten 2016). Here, the notable features are the many records during the Maunder minimum (1645 - 1715 CE) and the absence of records during 1771 - 1814 CE. As mentioned in section 1 that even without a sunspot, an eruption of large quiescent filament can cause geomagnetic storm (McAllister et al. 1996), the authors are not sure whether quiescent filament eruptions can produce such low-latitude auroras. Presumably, non-aurora events were considered, and are discussed in the next subsection. We did not have any clear interpretation for the absence of records during 1771 - 1814 CE. The other feature to be noted is that there is no record found corresponding to Carrington event (Carrington 1859). We will describe the statistical characteristics of records with the GSN in the next subsection.

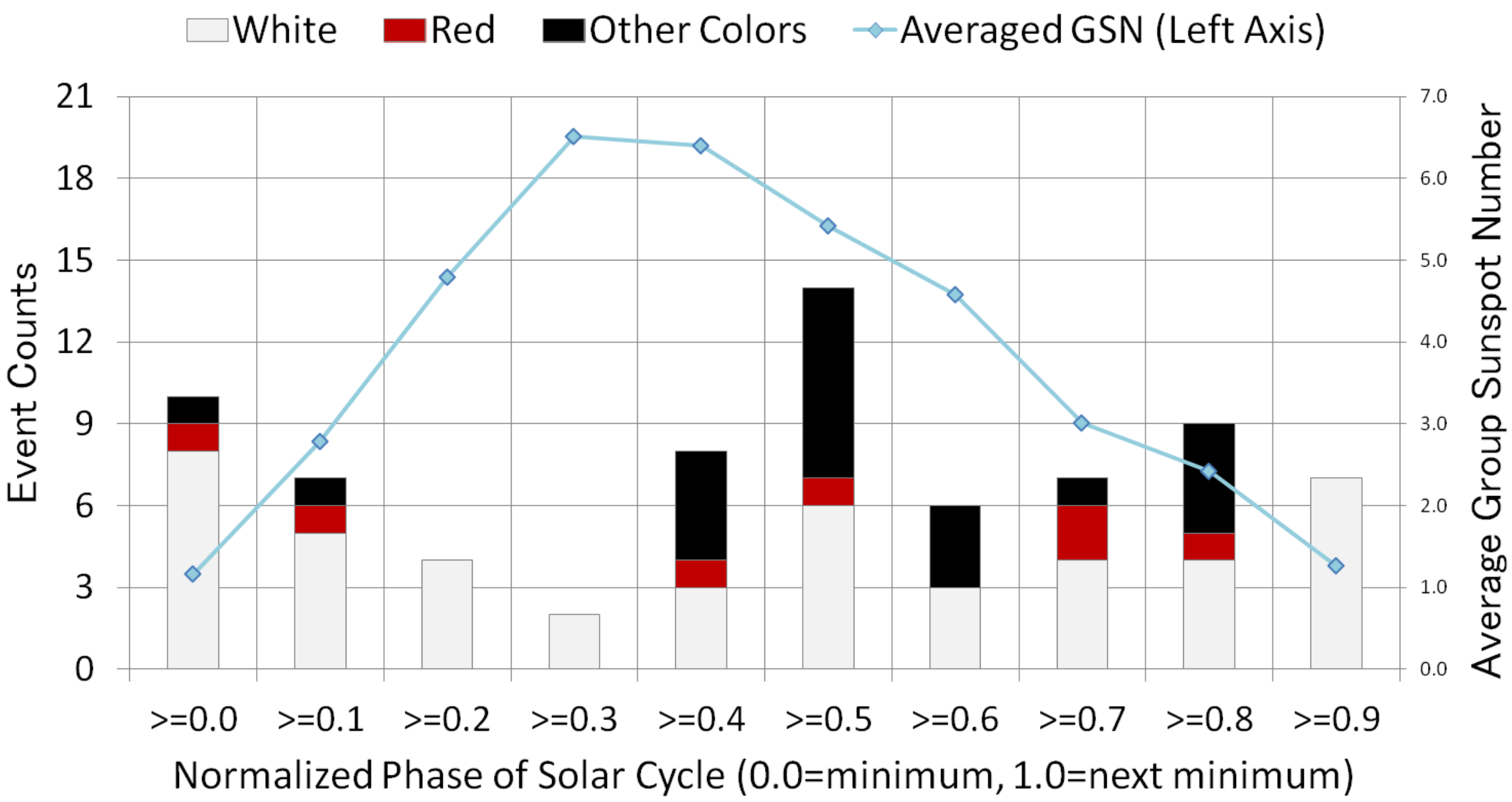

Figure 2. Histogram of event counts of white, red, or other colors (including multiple colors), painted in gray, red, and black, respectively, over the normalized phase of solar cycle with group sunspot number averaged over *Qīng* era for every 0.1 normalized phase of solar cycle. (color plot online)

### 3.4 Statistics over the Group Sunspot Number

A property of aurora statistics could be a relation to the GSN since the higher aurora activity is expected at the phase with more frequency of solar flare happening. From the study of geomagnetic disturbances (Kilpua et al. 2015), severe geomagnetic disturbances indicating high aurora activities tend to occur maximum and descending phase of a solar cycle. Our result plotted on the figure 2 shows that the majority of records of red or other colors are found in the phase between 0.4 and 0.9 of the normalized solar cycle. In contrast, considerable portion of the white records are found in the solar minimum (around the phase 0.0 or 1.0). Therefore, this result with red or other-color records does not contradict with the statistical property of geomagnetic disturbances, but white records may be contaminated with non-aurora phenomenon, such as some phenomena related to atmospheric optics.

### 3.5 Comparison with Western aurora records

Since auroras are global events, simultaneous observations of candidates at distant locations strongly indicate reliable aurora records. Fritz (1873) compiled a list of aurora (candidate) observations in Europe. The Fritz list includes over one thousand records, mostly from western and central Europe, and a smaller number of occurrences from various parts of Eurasia. The selection criterion of the source document is not clear.

We searched aurora records in the Fritz’s list that dated the same day or one day before the corresponding records in our list. We found 14 records in our list with correspondence to Fritz´s, which are listed in the following with our preliminary translations. In order to represent the ambiguity of original sentence, our preliminary translations contain some notations; “xx” denotes some missing or miswritten information, and translator’s comments are added with “[ ].” Also our translation leaves a question on interpretation of *chǐ* (尺) and *zhàng* (丈) which were usually used as units of length.

ID 67: 1730/02/15, lunar age=0.95:

夜分，福山見紅光滿野。

Translation: At midnight, at *Fúshān*, red light filling up the ground was observed.

ID 68: 1731/10/25, lunar age=0.83:

丑時，西北至東西，白雲二道，寬尺餘

Translation: Around 1:00 to 3:00, from northwest to eastwest [sic], two bands of white clouds, as wide as xx *chǐ*.

ID 72: 1739/09/23, lunar age=0.69:

北方白雲一道，寬尺餘，自東至西。

Translation: Northward one band of white cloud, as wide as xx *chǐ*, from east to west.

ID 80: 1747/12/23, lunar age=0.73:

卯時，東方白雲一道，寬尺餘，長丈餘。

Translation: Around 05:00 to 07:00 eastward, one band of white cloud, as wide as xx *chǐ*, as long as xx *zhàng*.

ID 81: 1753/09/27, lunar age=0.02:

東流有氣如虹著天，色紫白，久而沒。

Translation: At *Dōngliú*, there was vapor like a rainbow to reach the heaven with its color purple-white, and disappeared after a while.

ID 82: 1754/05/08, lunar age=0.54:

子時，中天白雲一道，自東南向西，寬尺餘，長二丈餘。

Translation: Around 23:00 to 01:00, at mid-heaven, one band of white cloud, from southeast to west, as wide as xx *chǐ*, as long as two *zhàng*.

ID 85: 1770/09/17, lunar age=0.94: (associated with events of 09/16-17 of Fritz 1873)

肥城有赤光自北方起，夜半漸退

Translation: At *Féichéng*, there was red vapor appeared from northward and gradually disappeared at midnight.

ID 86: 1770/09/17-18, lunar age=0.94: (associated with events of 09/16-17 of Fritz 1873)

長山西北見赤氣瀰天，中有白氣如縷間之，四更後始散。

Translation: At *Chángshān*, northwestward red vapor filled up the heaven, white vapor like thread in between it [red vapor], and after around 01:00 to 03:00 it started to disappear.

ID 87: 1770/09/17, lunar age=0.94: (associated with events of 09/16-17 of Fritz 1873)

肥城有白氣十三道，至夜半乃退。

Translation: At *Féichéng*, there were 13 bands of white vapors, which gradually disappeared until midnight.

ID 88: 1770/09/18, lunar age=0.98: (associated with events of 09/17 of Fritz 1873)

夜，榮成夜見紅光燭天。

Translation: At night, at *Róngchéng*, red vapor shining heaven like candles was observed.

ID 89: 1770/09/18, lunar age=0.98: (associated with events of 09/17 of Fritz 1873)

夜…東光有氣如火，橫蔽西北，亙數十丈，中含紅光，森如劍戟上射。

Translation: At *Dōngguāng*, there was vapor-like fire, covering northwest, as wide as several tens *zhàng*, including red light inside, as closely set as swords standing still.

ID 93: 1837/07/30, lunar age=0.94:

嵊縣有赤光如球，高數丈，三日乃滅。

Translation: There was red light like a ball, as high as several *zhàng*; after three days it disappeared.

ID 104: 1853/04/23, lunar age=0.51:

中衞有黑黃氣二道，直衝天際。

Translation: At *Zhōngwèi*, there were two bands of black-yellow vapors, directly hitting the border of the heaven.

ID 107: 1868/10/30, lunar age=0.50:

玉田有火光至空際化為白氣，長丈許，其中有聲如鼓。

Translation: At *Yùtián*, there was fire light arriving at the heaven and becoming white vapor, as long as xx *zhàng*, and in it [white vapor] there was a sound like drums.

The simultaneous observation in China and other parts of Eurasia strongly indicates that these were aurora observations, thus we considered these 14 records as primary candidates for auroras in the *Qīngshĭgăo*. Note that among the 14 primary candidates, five of them are considered as a part of the event of Sep. 17-18, 1770 CE.

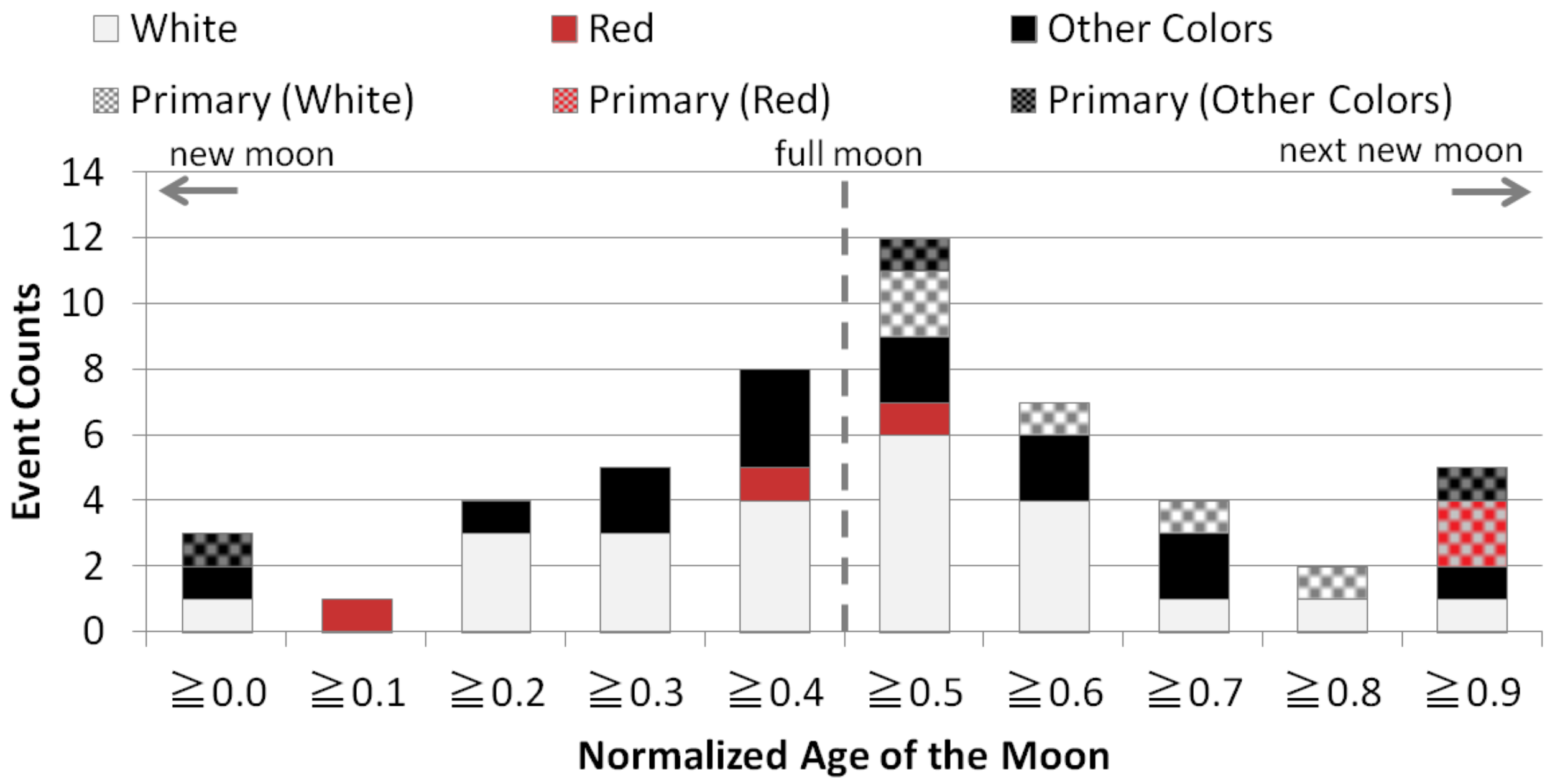

Figure 3. Distribution of aurora candidates from *Qīngshĭgăo* over the normalized moon age (0.0=new moon, 1.0=next new moon) with comet candidates excluded. Records of white, red, and other colors (including multiple colors) are painted in gray, red, and black respectively. In addition, primary candidates (painted in checker patterns) are presented as supplemental information. (Color online)

**3.6 Lunar age analysis**

As previously mentioned, the age of the moon is useful to examine the possibility of atmospheric optical events. For this purpose, we first excluded 26 records associated with comets (see subsection 3.3) and the records with incomplete date information, and then calculated the lunar age for the remaining 56 entries including the primary candidates of aurora. We made a record histogram concerning the normalized lunar age (0.0 for the new moon, 0.5 for around the full moon, and 1.0 for the next new moon) of every 0.1 normalized lunar ages, presented in Figure 3. It becomes clear that the distribution shows a peak near the full moon (around 0.5 normalized lunar age), particularly when excluding primary candidates. This trend is difficult to interpret merely with auroras because of their physical independence from lunar age.

The visibility of an aurora may be affected by the luminosity of the moon. Hence, the distribution of aurora observations over the lunar age will display a maximum around the new moon instead of the full moon, because of the moonlight-polluted night sky. This idea was also discussed by Vaquero et al. (2003), which associated the difficulty of naked-eye aurora observation with the moon appearance. On the other hand, moon-related atmospheric optical phenomena, such as 22° halo and moon dogs, are caused by both reflection and refraction of moon light in the air, mediated by hexagonal ice crystals. Due to its nature, this phenomenon requires a strong light source, like a full moon. Therefore, the trend involving a peak around the full moon could be reasonably interpreted in the framework of such atmospheric optical events. This does not mean that all the records reflected atmospheric optical events, except for the primary candidates, as they may include both auroras and atmospheric optical events.

Besides the primary candidates, there were five records around the new moon (0.0 - 0.1 and 0.9 - 1.0 normalized lunar age), which unlikely had lunar (atmospheric) origin. Therefore, we regarded these as secondary aurora candidates. These five records considered as secondary candidates are listed as following. Note that IDs 16 and 17 describe the event of same date.

ID 4: 1618/07/19, lunar age=0.93:

有紅、綠、白三氣，自天下垂，覆營左右，上圓如門。

Translation: There were three vapors whose colors were red, green, and white, suspended from the heaven, covering the encampment from left to right, as round as gates.

ID 16: 1650/06/30, lunar age=0.06:

泰安見白氣亙天

Translation: At *Tàiān* white vapor filling up the heaven was observed.

ID 17: 1650/06/30, lunar age=0.06:

益都見白氣亙天

Translation: At *Yìdōu* white vapor filling up the heaven was observed.

ID 26: 1667/10/14, lunar age=0.90:

有白光一道，自東至西。

Translation: There was one band of white light from east to west.

ID 47: 1679/09/07, lunar age=0.09:

正北黃黑雲一道，變赤黃色，寬四尺餘，長數丈。

Translation: Northward, one band of yellow-black cloud, changing its color to red-yellow, as wide as 4 *chǐ*, and as long as xx *zhàng*.

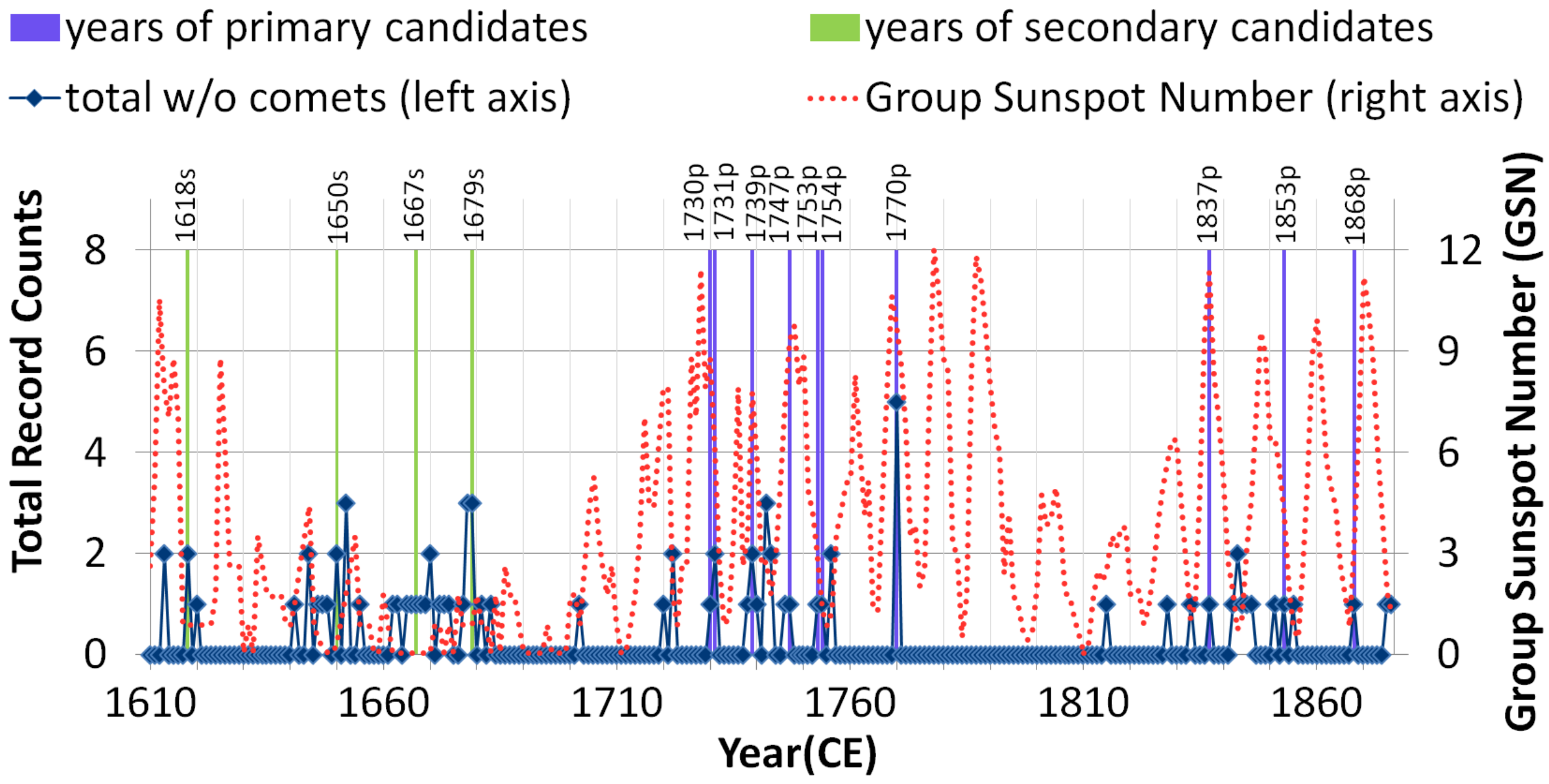

Figure 4. An update of figure1 with primary and secondary candidates, highlighted in purple and green, and labeled with “p” and “s”, respectively. (Color Online)

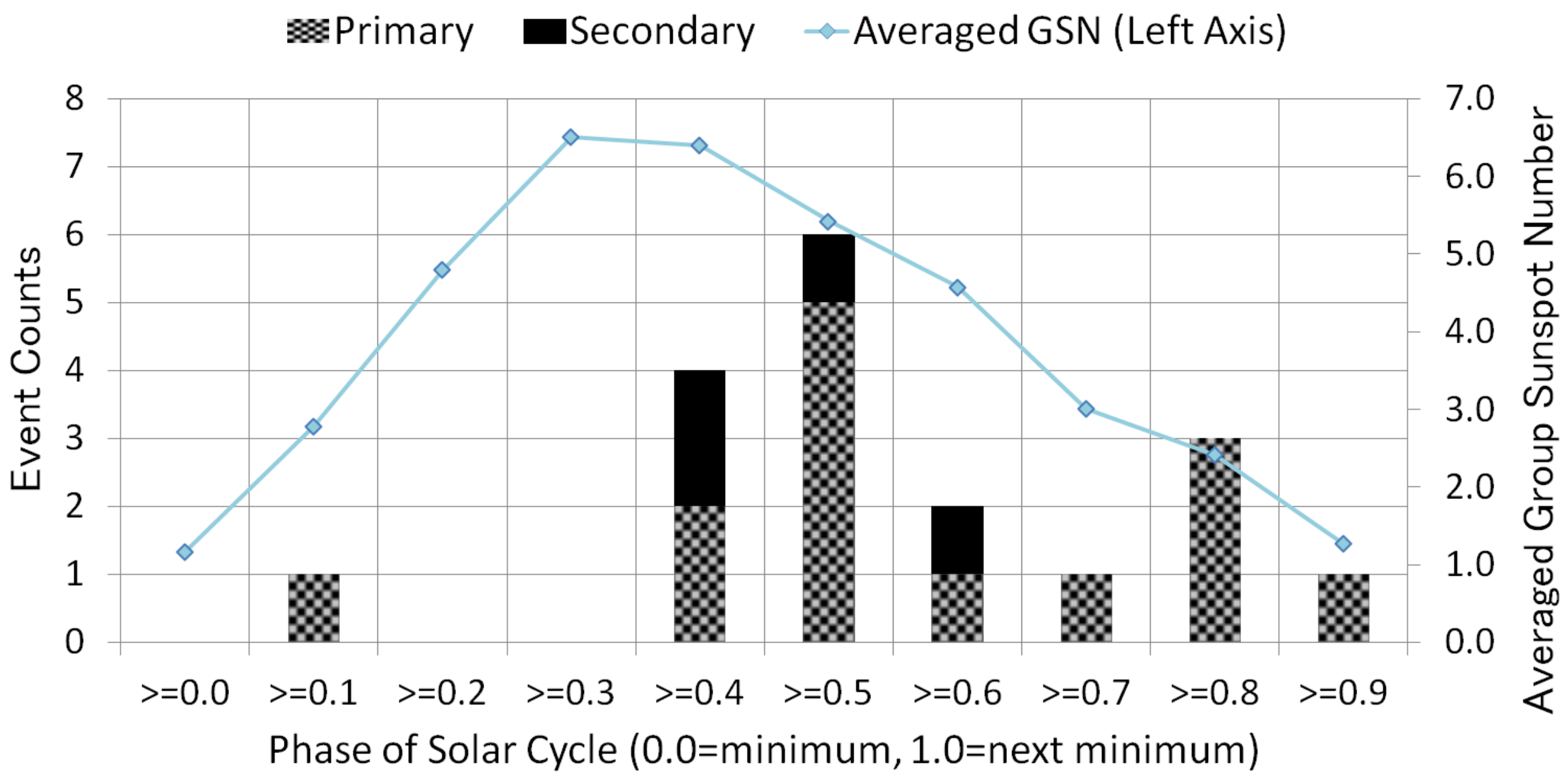

Figure 5. Similar to figure 2, but only wih primary and secondary candidates.

### 3.7 Discussions on primary and secondary candidates for aurora

At last, we examine the primary and secondary candidates with group sunspot numbers. The methods applied for this analysis on the primary and secondary candidates are same as subsections 3.2 and 3.4.

Figure 4, the updated version of figure 1, shows the timely occurrence of both primary and secondary aurora candidates, plotted with GSN and yearly-totaled record counts without comet candidates. The peak of 1770 is due to five records from multiple locations, which is very likely associated with the global appearance of aurora during 16 – 18 of Sep., 1770 CE at the local time of observation sites (Fritz 1870 for European observations). An interesting feature is that three events classified as secondary candidates occurred during the Maunder minimum; which events are of

1650/06/30 (IDs 16 and 17), 1667/10/14 (ID 26), and 1679/09/07 (ID 47).

Figure 5 provides the distribution of the primary and secondary candidates over the normalized solar cycle. Although both the primary and secondary candidates include events observed only in white, they are found to have happened after solar maximum. As we discussed in the section 3.4, this trend of the primary and secondary candidates does not conflict with the study on geomagnetic disturbances (Kilpua et al 2015).

## 4. Conclusion

As a result of our database survey of *Qīngshĭgăo* for searching aurora candidates during the *Qīng* era (1644-1912), we found 111 records of night-sky luminous events written neither as with the sun nor the moon, nor as during the daytime. However, 26 of 111 records are more likely understandable as comet observations. We classified 14 records as primary candidates for aurora, based on the simultaneous observations in other parts of Eurasia presented in the work by Fritz (1873). Five records were considered as secondary candidates for aurora according to lunar age analysis. Records of non-white colors (red, other colors, or multiple colors) are statistically agreeable with a study of geomagnetic disturbances. The remaining records especially of white are not completely excluded from aurora candidates, but the distributions over lunar age and the phase of a solar cycle suggest that a notable fraction of the records is probably related to atmospheric optical events. Our most significant finding during the coverage of *Qīngshĭgăo* is three secondary candidates during the Maunder minimum. We expect that a thorough analysis of these three events, combined with other perspectives, may encourage understandings of the solar activity at grand minimum.

**Acknowledgement**

Here we state a special thanks to Ken Saito for helping us identifying observation sites. We also acknowledge to the supports by the Center for the Promotion of Integrated Sciences (CPIS) of SOKENDAI as well as the Kyoto University's Supporting Program for Interaction-based Initiative Team Studies "Integrated study on human in space" (PI: H. Isobe), the Interdisciplinary Research Idea contest 2014 by the Center of Promotion Interdisciplinary Education and Research, the "UCHUCHUGAKU" project of the Unit of Synergetic Studies for Space, and the Exploratory Research Projects of the Research Institute of Sustainable Humanosphere, Kyoto University. This work was also encouraged by JSPS KAKENHI Grant Numbers 15H05816.

**References**

Carrington, R. C. 1859, Monthly Notices of the Royal Astronomical Society, 20, 13

Chen, H. Y. 2004, Historiography East and West, 2, 2, 173

Clette, F., Svalgaard, L., Vaquero, J. M., & Cliver, E. W. 2014, Space Science Reviews, 186, 1-4, 35

Cliver, E.W. & Ling A.G. 2016, Solar Physics, DOI: 10.1007/s11207-015-0841-6

Basurah, H. 2006, Journal of Atmospheric and Solar-Terrestrial Physics, 68, 937

Broughton, P. 2002, J. Geophys. Res., Space Phys. 107, A8, SIA 1

Dall'Olmo, U. 1979, J. Geophys. Res., Space Phys., 84, 1525

Eddy, J. A. 1980, in Proc. The ancient sun: Fossil record in the earth, moon and meteorites, 119

Fritz, H. 1873, Verzeichniss Beobachteter Polarlichter, C. Gerold's & Sohn, Wien

Hayakawa, H., Tamazawa, H., Kawamura, A. D., & Isobe, H. 2015, Earth, Planets and Space, 67, 82

Hoyt, D. V., & Schatten, K. H. 1998a, Solar Physics, 179, 189-219

Hoyt, D. V., & Schatten, K. H. 1998b, Solar Physics, 181, 491-512

Kanda, S. 1933, Astron Herald, 26, 11, 204 (in Japanese)

Keimatsu, M. 1970, Annals of Science of Kanazawa University 7, 1

Keimatsu, M. 1971, Annals of Science of Kanazawa University 8, 1

Keimatsu, M. 1972, Annals of Science of Kanazawa University 9, 1

Keimatsu, M. 1973, Annals of Science of Kanazawa University 10, 1

Keimatsu, M. 1974, Annals of Science of Kanazawa University 11, 1

Keimatsu, M. 1975, Annals of Science of Kanazawa University 12, 1

Keimatsu, M. 1976, Annals of Science of Kanazawa University 13, 1

Kilpua, E. K. J., et al. 2015, ApJ. 806, 272

Lee, E., Ahn, Y., Yang, H., & Chen, K. 2004, Sol. Phys., 224, 373

Link, F. 1962, Geofysikální Sborník, 173, 297

Lynn, W. T. 1888, The Observatory, 11, 437

Maehara, H., et al. 2012, Nature, 485, 478

Matsushita, S. 1956, J. Geophys. Res., 61, 297

Meeus, J. 1998, Astronomical algorithms, Willmann-Bell, Inc., Richmond

McAllister, A. H., Dryer, M., Mclntosh, P., Singer, H., & Weiss, L. 1996, J. Geophys. Res., Space phys., 101, A6, 13497

Miyake, F., Nagaya, K., Masuda, K., & Nakamura, T. 2012, Nature, 486, 240

Miyake, F., Masuda, K., & Nakamura, T. 2013, Nature Com., 4, 1748

Miyahara, H., Masuda, K., Muraki, Y., Furuzawa, H., Menjo, H., & Nakamura, T. 2004, Solar Physics, 224, 317

Miyahara, H., Yokoyama, Y., & Kimiaki, M. 2008, Earth and Planetary Science Letters, 272, 290

Miyahara, H 2010, Journal of Geography, 119, 3, 510

Nakazawa, Y., Okada, T., & Shiokawa, K. 2004, Earth, Planets, and Space, 56, e41

Notsu, Y., Honda, S., Maehara, H., Notsu, S., Shibayama, T., Nogami, D., & Shibata, K. 2015a, PASJ, 67, 32

Notsu, Y., Honda, S., Maehara, H., Notsu, S., Shibayama, T., Nogami, D., & Shibata, K. 2015b, PASJ, 67, 33

Saito, K., & Ozawa, K. 1992, Examination of Chinese ancient astronomical records (中国古代の天文記録の検証) . Yuzankaku Press, Tokyo (in Japanese)

Schaefer, B. E., King, J. R., & Deliyannis, C. P. 2000, ApJ, 529, 1026

Schrijver, C. J. et al. 2012 J. Geophys. Res., Space Phys., 117, A08103

Schove, D. J., & Ho, P.-Y. 1967, Journal of the American Oriental Society, 87, 2, 105

Shibayama, T. et al. 2013, ApJS, 209, 5

Shiokawa, K., Ogawa, T., & Kamide, Y. 2005, J. Geophys. Res., 110, 9, A05202

Stothers, R. 1979, Isis, 70, 1, 85

Svalgaard, L., & Schatten, K. H. 2016, Solar Physics, DOI: 10.1007/s11207-015-0815-8

Usoskin, I.G., Kovaltsov, G.A., Lockwood, M., Mursula, K., Owens, M., & Solanki, S.K. 2016, Solar Physics, DOI: 10.1007/s11207-015-0838-1

Vaquero, J. M., & Galleg, M. C. 2001, Annales Geophysicae, 19, 809

Vaquero, J.M., Gallego M.C., & García, J. A. 2003, Journal of Atmospheric and Solar-Terrestrial Physics, 65, 6, 677

Vaquero, J. M., & Trigo, R. M. 2005, Solar Phys., 231, 157

Vaquero, J. M., Gallego, M. C., Barriendos, M., Rama, E., & Sánchez-Lorenzo, A. 2010, Advances in Space Research, 45, 11, 1388

Vázquez, M., & Vaquero, J. M. 2010, Sol. Phys., 267, 431

Vyssotsky, A. N. 1949, Meddelande Fran Lunds Astronomiska Observatorium, Ser. II, No. 126, Historical papers, No 22

Yau, K. K. C., & Stephenson, F. R. 1988, Quarterly Journal of Royal Astronomical Society, 29, 175

Yau, K. K. C., Stephenson, F. R., & Willis, D.M. 1995, Rutherford Appleton Lab Technical Reports,

RAL-TR-95-073

Yeomans, D. K. 1991 Comets. A chronological history of observation, science, myth, and folklore, Wiley, New York

Zhào, Ě. 1976, *Qīngshĭgăo*, *Bĕijīng*: *Zhōnghuá Shūjú*

## Table 1 List of Aurora Candidates during the Qīng Dynasty

Treasures of the List of Aurora Candidates during the *Qīng* Dynasty

Color: B - blue, Bl - black, G - green, Gl - Gold, R - red, W - white, Y - yellow

Description: V - vapor (氣, *qì*), C - cloud (雲, *yún*), L - light (光, *guāng*)

Direction: e - east, n - north, s - south, w - west

Length: c - *chǐ* (尺), z - *zhàng* (丈), if both width and length are given then noted as “width*length.”

Place: See the Appendix.B

Lunar age: 0.0 (new moon), 1.0 (next new moon), blank (no valid date information)

Refer Table 2 for the list of observation places

More detailed version of this table is available at (http://www.kwasan.kyoto-u.ac.jp/~palaeo/).

| ID | Year | Mo nth | Day | Hour(0-23, 2hour system) | Durati on | Durati on Unit | Color | Descript ion | Directio n | Length | counts | Place | Notes | Lunar Phase |
|---|---|---|---|---|---|---|---|---|---|---|---|---|---|---|
| 1 | 1613 | 1 | | | | | BW | V | | | | | | |
| 2 | 1613 | 1 | | | | | W | V | | | | | | |
| 3 | 1618 | 2 | 10 | | | | Y | V | | 2c*3-4z | | | penetrating moon | 0.52 |
| 4 | 1618 | 7 | 19 | | | | R, G, W | V | | | 3 | | like a gate | 0.93 |
| 5 | 1618 | 11 | 15 | | 16 | days | W | V | es | 15z | | | | 0.95 |
| 6 | 1620 | 4 | | | | | B, W | V | w-e | | | | around moon | |
| 7 | 1641 | 10 | 12 | | | | Gl | L | e | | | | at a sunrise | 0.27 |
| 8 | 1644 | 7 | 17 | 18 | | | W | V | ws-en | | | | 17:00-19:00 | 0.46 |
| 9 | 1644 | 3 | | | | | W | V | | | 1 | | | |
| 10 | 1646 | 3 | 1 | | | | R | L | n | | | | like a shade of fire | 0.48 |
| 11 | 1647 | 6 | 12 | | | | W | V | ws-en | | | | | 0.33 |
| 12 | 1648 | 5~7 | | | | | R | V | e | | | *Jiādìng* | | |
| 13 | 1649 | 4 | | | 1 | month | W | V | | | | *Jiāngyīn* | | |
| 14 | 1650 | 2 | 26 | | 10 | days | W | V | w | | | *Kūnshān* | | 0.88 |
| 15 | 1650 | 2 | 26 | | 20+ | days | W | V | w | | | *Xiāoxiàn* | | 0.88 |
| 16 | 1650 | 6 | 30 | | | | W | V | | | | *Tàiān* | | 0.06 |
| 17 | 1650 | 6 | 30 | | | | W | V | | | | *Yìdū* | | 0.06 |
| 18 | 1652 | 1 | 19 | | | | W | V | en-wn | | | | | 0.29 |
| 19 | 1652 | 2 | | | | | R | L | e-es | | | *Dōngchāng* | sound | |
| 20 | 1652 | 12 | 22 | | | | BW | V | | | | | | 0.73 |
| 21 | 1655 | 7 | 20 | | | | BBl | CV | n | | | | like a dragon | 0.58 |
| 22 | 1662 | 1 | 31 | | | | W | V | | | | *Qīxiá* | | 0.39 |
| 23 | 1663 | | | | | | W | V | | | | *Láiyáng* | | |
| 24 | 1665 | 2 | 1 | | | | W | V | | 3z | 1 | | around venus | 0.55 |
| 25 | 1666 | 3 | 14 | | | | BW | V | | | 4~5 | | | 0.30 |
| 26 | 1667 | 10 | 14 | | | | W | L | e-w | | 1 | | | 0.91 |
| 27 | 1668 | 2 | | | 20+ | days | W | V | w-e | | | *Guǎngpíng* | | |
| 28 | 1668 | 2 | | | 5-6 | days | W | V | | | | *Nèiqiū* | | |
| 29 | 1668 | 2 | | | 4 | days | W | V | w | some tens z | | | | |
| 30 | 1668 | 2 | | | | | W | V | w-e | | | *Wēixiàn* | | |
| 31 | 1668 | 3 | 7 | | 10+ | days | W | L | es | 6z-4z | 1 | | | 0.83 |
| 32 | 1668 | 3 | | | 40+ | days | W | V | | 10+z | | *Guǎngzhōu* | like a spear | |
| 33 | 1668 | 3 | | | | | W | V | | | | *Wǔyì* | | |
| 34 | 1668 | 3 | | | | | W | V | | | | *Tángshān* | | |
| 35 | 1668 | 8 | | | | | W | V | w | | | *Gāoyì* | | |
| 36 | 1669 | 7 | 20 | | | | | V | wn | | 1 | | | 0.76 |
| 37 | 1670 | 4 | 27 | | | | W | V | w | | | *Lúlíng* | | 0.27 |
| 38 | 1670 | 12 | | | | | W | V | s-n | | | *Tōngwèi* | like a rainbow | |
| 39 | 1672 | 8 | 6 | | | | W | V | ws-en | | | *Jiāohé* | sound | 0.46 |
| 40 | 1673 | 3 | 7 | | | | BW | V | wn-es | | | | like a textile | 0.64 |
| 41 | 1674 | | | | | | BW | C | en | | 1 | | | |
| 42 | 1677 | 8 | | | | | W | V | e-w? | | | *Lúlóng* | like a rainbow | |
| 43 | 1678 | 7 | 30 | | | | B | V | | 5c | 1 | | | 0.41 |
| 44 | 1678 | 7 | 31 | | | | BW,B | V | | | 1, 3 | | | 0.44 |
| 45 | 1678 | 8 | 1 | | | | B | V | wn-es | 6c | | | | 0.48 |
| 46 | 1679 | 7 | 31 | | | | W | V | | | | *Wǔdìng* | | 0.81 |
| 47 | 1679 | 9 | 7 | | | | YBl | C | | 4c*some z | 1 | | | 0.09 |
| 48 | 1679 | 12 | | | | | W | V | ws | | | *Hànzhōng* | | |
| 49 | 1680 | 11 | | | 1 | month | W | V | w | | | *Quánjiāo* | | |
| 50 | 1680 | 12 | 21 | | | | W | V | ws-en | | | *Cāngzhōu* | like a broom | 0.03 |
| 51 | 1680 | 12 | 21 | | | | W | V | e | | | *Lúlóng* | like a cloud | 0.03 |
| 52 | 1680 | 12 | 21 | | | | W | V | | | | *Jiàngxiàn* | like a rainbow | 0.03 |
| 53 | 1680 | 12 | 22 | | | | W | V | w | | | *Zhènyáng* | | 0.06 |
| 54 | 1680 | 12 | 22 | | | | W | V | w-e | | | *Línzī* | | 0.06 |
| 55 | 1680 | 12 | 23 | | | | BW | V | ws-en | | 1 | | | 0.09 |
| 56 | 1680 | 12 | 24 | | 1 | month | W | V | | 10+z | | *Wēnzhōu* | like a silk textile | 0.13 |
| 57 | 1681 | 7 | 24 | | | | B | V | | | 6 | | | 0.32 |
| 58 | 1681 | 8 | 4 | | 48 | days | W | V | | | | *Wàngjiāng* | | 0.70 |
| 59 | 1681 | 11 | | | 1 | month | W | V | | | | *Shānyáng* | | |
| 60 | 1681 | 11 | | | | | W | V | | | | *Hànzhōng* | like a silk textile | |
| 61 | 1683 | 6 | 12 | | | | W | V | | | | *Qīnghé* | like a rainbow | 0.60 |
| 62 | 1700 | 10 | | | 6-7 | days | W | V | | | | *Jiāngxià* | like silk | |
| 63 | 1702 | 3 | | | | | W | V | w | | | *Pèixiàn* | | |
| 64 | 1720 | 8 | 19 | | | | R | V | e-wn | | | *Róngchéng* | sound | 0.53 |
| 65 | 1722 | 1 | 6 | | | | W | V | ws | | | *Zūnhuà* | | 0.64 |
| 66 | 1722 | 7 | 26 | | | | W | V | | | | *Jiādìng* | | 0.45 |
| 67 | 1730 | 2 | 15 | | | | R | L | | | | *Fúshān* | | 0.95 |
| 68 | 1731 | 10 | 25 | | | | W | C | wn-es | | 2 | | | 0.83 |
| 69 | 1731 | | | | | | W | V | | | 1 | *Nángōng* | sound (error dated) | |
| 70 | 1738 | 9 | 2 | | | | W | C | wn-ws | 3c*1z | | | | 0.62 |
| 71 | 1739 | 4 | 26 | | | | | C | es | | 1 | | | 0.63 |
| 72 | 1739 | 9 | 23 | | | | W | C | e-w | | 1 | | | 0.70 |
| 73 | 1740 | 4 | 12 | | | | W | C | es | 3z | 1 | | | 0.55 |
| 74 | 1742 | 2 | 22 | 0 | | | W | C | | 3z | 1 | | 23:00-01:00, under moon | 0.59 |
| 75 | 1742 | 3 | 24 | | | | W | C | n | 2c | 1 | | | 0.61 |
| 76 | 1742 | 9 | 21 | 0 | | | W | C | e | | 1 | | 23:00-01:00 | 0.77 |
| 77 | 1743 | 5 | 31 | 0 | | | W | C | | | 1 | | 23:00-01:00, above moon | 0.28 |
| 78 | 1743 | 9 | 5 | | | | W | C | e-w | | | | | 0.58 |
| 79 | 1746 | 8 | 27 | | | | W | C | | | 1 | | | 0.37 |
| 80 | 1747 | 12 | 23 | | | | W | C | e | | 1 | | | 0.73 |
| 81 | 1753 | 9 | 27 | | | | PW | V | | | | *Dōngliú* | like a rainbow | 0.03 |
| 82 | 1754 | 5 | 8 | | | | W | C | es-w | 2z | 1 | | | 0.55 |
| 83 | 1756 | 6 | 11 | | | | W | C | es | | 1 | | | 0.47 |
| 84 | 1756 | 11 | 12 | 2 | | | W | C | | | 1 | | 01:00-03:00 | 0.68 |
| 85 | 1770 | 9 | 17 | | ? | hours | R | L | n | | | *Féichéng* | until midnight | 0.95 |
| 86 | 1770 | 9 | 17 | | | | R,W | V | wn | | | *Chángshān* | | 0.95 |
| 87 | 1770 | 9 | 17 | | ? | hours | W | V | | | 13 | *Féichéng* | until midnight | 0.95 |
| 88 | 1770 | 9 | 18 | | | | R | L | | | | *Róngchéng* | | 0.98 |
| 89 | 1770 | 9 | 18 | | | | R | L | wn | some tens z | | *Dōngguāng* | | 0.98 |
| 90 | 1815 | 6 | | | | | W | V | w | some z | | *Wǔdìng* | | |
| 91 | 1828 | 1 | | | | | W | V | | | some tens | *Xiāoxiàn* | | |
| 92 | 1833 | 6 | 5 | | | | W | V | | | | *Qīxiá* | | 0.59 |
| 93 | 1837 | 7 | 30 | | | | R | L | | some z | | *Shèngxiàn* | like a ball | 0.94 |
| 94 | 1840 | | | | 1 | month | W | V | | | | *Chānglí* | | |
| 95 | 1842 | | | | 1 | month | W | V | | some z | | *Shēnxiàn* | like a textile | |
| 96 | 1842 | | | | | | W | V | | | | *Hànzhōng* | | |
| 97 | 1843 | 5 | | | | | W | V | ws | | | *Huángzhōu* | like a textile | |
| 98 | 1843 | 6 | | | 1 | month | W | V | | | | *Téngxiàn* | | |
| 99 | 1843 | | | | | | W | V | ws | | | | | |
| 100 | 1844 | | | | | | W | V | | | | *Dēngzhōu* | | |
| 101 | 1845 | | | | | | W | V | wn | | | *Jímò* | | |
| 102 | 1846 | | | | | | W | L | | | | *Níngjīn* | | |
| 103 | 1851 | | | | | | W | L | | | | *Huángān* | sound | |
| 104 | 1853 | 4 | 23 | | | | BlY | V | | | 2 | | | 0.52 |
| 105 | 1855 | 9 | | | | | R | V | e | | | *Cáoxiàn* | like a flag | |
| 106 | 1862 | | | | 10 | days | W | L | wn | | | *Qīxiá* | | |
| 107 | 1868 | 10 | 30 | | | | W | V | | | | *Hànzhōng* | sound | 0.50 |
| 108 | 1875 | | | | | | W | V | | | | *Jiādìng* | | |
| 109 | 1876 | 2 | 28 | | | | R | L | | | | *Cáoxiàn* | | 0.12 |
| 110 | | | | | | | B, W | V | e | | 2 | | | |
| 111 | | | | | | | Y | V | | | | | | |

**Table 2**

53 observation cities

| City (Chinese letter) | City (Pinyin) | Latitude (North) | Longitude (East) |
|---|---|---|---|
| 滄州 | *Cāngzhōu* | 38°18′ | 116°50′ |
| 曹縣 | *Cáoxiàn* | 34°49′ | 115°32′ |
| 昌黎 | *Chānglí* | 39°42′ | 119°9′ |
| 長山 | *Chángshān* | 36°52′ | 117°52′ |
| 登州 | *Dēngzhōu* | 37°48′ | 120°45′ |
| 東昌 | *Dōngchāng* | 36°27′ | 115°59′ |
| 東光 | *Dōngguāng* | 37°53′ | 116°32′ |
| 東流 | *Dōngliú* | 30°12′ | 116°54′ |
| 肥城 | *Féichéng* | 36°10′ | 116°46′ |
| 福山 | *Fúshān* | 37°29′ | 121°16′ |
| 高邑 | *Gāoyì* | 37°36′ | 114°36′ |
| 廣平 | *Guăngpíng* | 36°29′ | 114°56′ |
| 廣州 | *Guăngzhōu* | 23°7′ | 113°15′ |
| 海陽 | *Hăiyáng* | 36°46′ | 121°9′ |
| 漢中 | *Hànzhōng* | 33°4′ | 107°1′ |
| 黃安 | *Huángān* | 31°17′ | 114°37′ |
| 黃州 | *Huángzhōu* | 30°26′ | 114°52′ |
| 即墨 | *Jímò* | 36°23′ | 120°26′ |
| 嘉定 | *Jiādìng* | 29°33′ | 103°45′ |
| 江夏 | *Jiāngxià* | 30°33′ | 114°18′ |
| 絳縣 | *Jiàngxiàn* | 35°29′ | 111°34′ |
| 江陰 | *Jiāngyīn* | 31°55′ | 120°17′ |
| 交河 | *Jiāohé* | 42°57′ | 89°3′ |
| 崑山 | *Kūnshān* | 31°23′ | 120°58′ |
| 萊陽 | *Láiyáng* | 36°58′ | 120°42′ |
| 臨淄 | *Línzī* | 36°49′ | 118°18′ |
| 廬陵 | *Lúlíng* | 27°6′ | 114°59′ |
| 盧龍 | *Lúlóng* | 39°59′ / 40°25′ | 118°54′ / 118°20′ |
| 南宮 | *Nángōng* | 37°21′ | 115°24′ |
| 内丘 | *Nèiqiū* | 37°17′ | 114°30′ |
| 寧津 | *Níngjīn* | 37°39′ | 116°48′ |
| 沛縣 | *Pèixiàn* | 34°45′ | 116°56′ |
| 棲霞 | *Qīxiá / Xīxiá* | 37°20′ | 120°50′ |
| 清河 | *Qīnghé* | 33°37′ | 119°2′ |
| 全椒 | *Quánjiāo* | 32°5′ | 118°16′ |
| 榮成 | *Róngchéng* | 37°7′ | 122°29′ |
| 山陽 | *Shānyáng* | 33°36′ | 119°0′ |
| 莘縣 | *Shēnxiàn* | 36°14′ | 115°40′ |
| 嵊縣 | *Shèngxiàn* | 29°33′ | 120°50′ |
| 泰安 | *Tàiān* | 36°12′ | 117°5′ |
| 唐山 | *Tángshān* | 39°37′ | 118°10′ |
| 滕縣 | *Téngxiàn* | 35°5′ | 117°9′ |
| 通渭 | *Tōngwèi* | 35°12′ | 105°14′ |
| 望江 | *Wàngjiāng* | 30°4′ | 116°48′ |
| 威縣 | *Wēixiàn* | 36°58′ | 115°16′ |
| 溫州 | *Wēnzhōu* | 27°59′ | 120°41′ |
| 武定 | *Wŭdìng* | 25°31′ | 102°24′ |
| 武邑 | *Wŭyì* | 37°48′ | 115°53′ |
| 蕭縣 | *Xiāoxiàn* | 34°11′ | 116°56′ |
| 益都 | *Yìdū* | 36°41′ | 118°28′ |
| 玉田 | *Yùtián* | 39°54′ | 117°44′ |
| 鎮洋 | *Zhènyáng* | 31°27′ | 121°7′ |
| 遵化 | *Zūnhuà* | 40°11′ | 117°57′ |